\documentclass[letterpaper,10pt]{article} 

\usepackage{osameet3} 

\usepackage{amsmath,amssymb, amsfonts, amsthm, bm}
\usepackage{newtxmath}
\usepackage{commath}
\usepackage[colorlinks=true,bookmarks=false,citecolor=blue,urlcolor=blue]{hyperref} 

\usepackage{lipsum}
\usepackage[mode=tex]{standalone} 

\usepackage{subfigure}
\usepackage{wrapfig}

\usepackage{dsfont}
\usepackage{float}
\usepackage{algorithm}
\usepackage[noend]{algpseudocode}
\usepackage{tikz,pgfplots}
\usepgfplotslibrary{fillbetween}
\usepgfplotslibrary{groupplots}
\usetikzlibrary{shapes.geometric}

\usepackage{paralist}
\usepackage{xcolor}
\tikzset{%
	partial ellipse/.style args={#1:#2:#3}{%
		insert path={+ (#1:#3) arc (#1:#2:#3)}%
	}%
}%

\makeatletter
\def\BState{\State\hskip-\ALG@thistlm}
\makeatother

\newcommand{\imag}{j}

\newcommand{\mysubsection}[1]{%
    \emph{#1}: 
}%

\begin{document}
\title{FPGA Implementation of Multi-Layer Machine Learning Equalizer with On-Chip Training}

\vspace{-0.6cm}
\author{
    Keren Liu\textsuperscript{(1)},  
    Erik Börjeson\textsuperscript{(1)},  
    Christian H\"{a}ger\textsuperscript{(2)}, and
    Per Larsson-Edefors\textsuperscript{(1)}
}
\address{\textsuperscript{(1)} Department of Computer Science and Engineering, Chalmers University of Technology, Gothenburg, Sweden\\
   \textsuperscript{(2)} Department of Electrical Engineering, Chalmers University of Technology, Gothenburg, Sweden\\
    }
\email{christian.haeger@chalmers.se, perla@chalmers.se}

\copyrightyear{2023}
\vspace{-0.5cm}
\begin{abstract}
We design and implement an adaptive machine learning equalizer that alternates multiple linear and nonlinear computational layers on an FPGA. 
On-chip training via gradient backpropagation is shown to allow for real-time adaptation to time-varying channel impairments.  
\end{abstract}

\vspace{-0.03cm}

\section{Introduction}

Optical fiber channels suffer from both linear and nonlinear impairments that severely affect the transmission performance. 
Moreover, environmental changes due to temperature or mechanical strains can lead to time-varying effects which require adaptive equalization. 
Adaptive equalizers are indeed commonplace in optical receivers \cite{Chen2012, Crivelli2014}, typically implemented via gradient-descent-based least-mean squares filtering \cite{Diniz2013}. 
For example, in coherent systems such equalizers can track the inverse Jones matrix of the channel and may also correct for additional distortions such as residual chromatic dispersion \cite{Savory2008}.
However, the underlying equalizer structure is \emph{linear}, which limits the type of functionalities that can be expressed and therefore also the performance that can be achieved.

To overcome the limitations of linear equalizers, a wide variety of machine learning (ML) algorithms have recently been proposed and verified in hardware (HW).
For example, field-programmable gate array (FPGA) implementations of various neural network equalizers were demonstrated for IM/DD links \cite{Chagnon2019}, passive optical networks \cite{Kaneda2020}, optical interconnects \cite{Li2021}, and coherent systems \cite{Freire2022ecoc}. 
Moreover, application-specific integrated circuit (ASIC) design of a model-based ML equalizer \cite{Haeger2018ofc, Haeger2021jsac} was studied in \cite{Fougstedt2018ecoc}. 
However, all of the previous works in \cite{Chagnon2019, Kaneda2020, Li2021, Freire2022ecoc, Fougstedt2018ecoc} consider \emph{static} nonlinear equalization, i.e., the training is performed offline and only the inference stage is implemented in HW. 
By contrast, in this paper we implement \emph{both} the inference and training stage of a model-based ML equalizer on the same FPGA, which allows the equalizer to adapt to time-varying channel impairments. 
To the best of our knowledge, this is the first paper that studies HW implementation of on-chip gradient backpropagation \cite{Rumelhart1986} for nonlinear equalizers.
{Note that an adaptive equalizer based on unsupervised $K$-means clustering was implemented on an FPGA in \cite{Giacoumidis2020}. 
However, the corresponding training stage is different (and less complex) compared to the gradient-based training of neural networks.}

\section{Machine Learning Equalizer Model}

\begin{figure}[b]
    \vspace{-0.3cm}
    \centering
    \includegraphics[width=0.88\columnwidth]{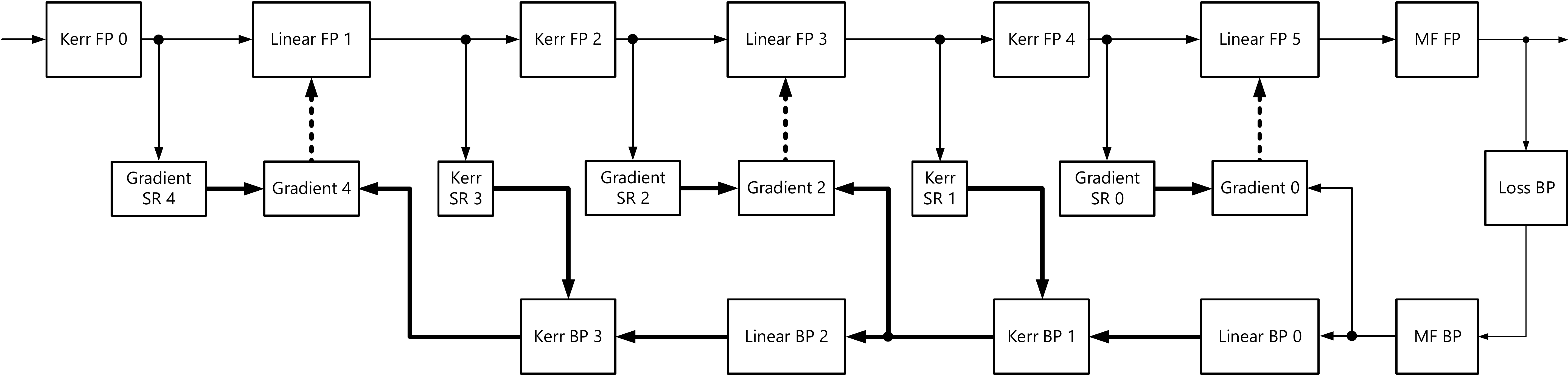}
    \vspace{-0.3cm}
    \caption{Equalizer structure including forward propagation (FP) and gradient backward propagation (BP). (SR: shift register)}
    \vspace{-0.6cm}
    \label{fig:equalizer}
\end{figure}

Following \cite{ModelBased}, our ML equalizer is based on the split-step solution of the (inverse) Manakov-PMD equation. 
The equalizer (Fig.~\ref{fig:equalizer}) consists of 3 layers and includes both the forward propagation (FP) and gradient backward propagation (BP). 
The FP alternates trainable linear steps and fixed (non-trainable) nonlinear Kerr steps. 
Each linear step applies a real-valued $2\times 2$ multiple-input multiple-output finite impulse response (MIMO-FIR) filter independently to the real and imaginary parts of the signal in both polarizations. 
Each nonlinear step applies $\boldsymbol{u} \exp(\imag \frac{8}{9} \gamma L \|\boldsymbol{u}\|^2  )$, where $\boldsymbol{u} \in \mathbb{C}^2$ is the Jones vector of the signal, $\gamma$ the nonlinearity parameter, and $L$ the step size.
As a last step, a non-trainable matched filter (MF) is applied independently to the signal in each polarization. 

The equalizer is trained via stochastic gradient descent according to $\theta_{i+1} = \theta_i - \xi \nabla \mathcal{L}(\theta_i)$, where $\theta_i$ are the trainable parameters in iteration $i$, $\xi$ is the learning rate, and $\mathcal{L}$ is a loss function. 
We use the mean squared error (MSE) $\mathcal{L}(\theta_i) = \frac{1}{B} \sum_{k=1}^B \|\boldsymbol{y}_k(\theta_i) - \boldsymbol{x}_k\|^2$, where $B$ is the batch size, $\boldsymbol{y}_k(\theta_i) \in \mathbb{C}^2$ are the estimated symbols after the MF, and $\boldsymbol{x}_k \in \mathbb{C}^2$ are known pilots. 
This corresponds to standard supervised training. 
However, different loss functions can also accommodate blind or decision-directed training modes without major HW modifications, similar to conventional adaptive linear equalizers \cite{Crivelli2014}. 
The gradients $\nabla \mathcal{L}(\theta_i)$ are calculated numerically by applying the chain rule \cite{Rumelhart1986}. 
The data flow in the BP is reversed compared to the FP (see the bottom of Fig.~\ref{fig:equalizer}), where the calculation of local derivatives require as input the derivatives from the previous layer and intermediate signals from the FP. 

\section{Hardware (HW) Implementation}

After our implementations are developed and verified at the system level, using TensorFlow and  Matlab, we develop HW description code (VHDL) to logically and sequentially define the HW structure. 
Final HW verification and analysis is performed with logic simulation: The Matlab system model (see Sec.~\ref{sec:results}) supplies the equalizer's VHDL model with input data cycle by cycle and the VHDL output is verified with system-level reference data. 

\mysubsection{Forward Propagation (Inference)}
Running at 2 samples/symbol, the FP HW uses short time-domain filters~\cite{ModelBased,Fougstedt2017ofc} and a HW-friendly Kerr layer~\cite{Liu2022sppcom} to simplify implementation. Since all HW is pipelined for throughput, the FP HW needs to use delays implemented as shift registers (see Fig.~\ref{fig:equalizer}) to synchronize the FP and BP inputs. 

\mysubsection{Backward Propagation and Training}
BP was first implemented by hand in Matlab and verified against the automatic gradient computation in TensorFlow. Next, using the Matlab implementation as a reference, we built a structural HW model of the equalizer using VHDL. 
Since the complexity of BP HW is potentially very high, this would have negative consequences on HW resource usage and on cycle latency, which impacts the timing with which the BP updates the equalizer parameters. 
To have an efficient HW implementation, we used several different techniques across the BP modules: (i) Addition of several fixed right shifts $\frac{1}{2^n}$ to perform division by the batch size $B$ in the loss BP layer, (ii) application of HW-efficient Taylor expansion~\cite{Fougstedt2017ofc, Fougstedt2018ecoc} to the Kerr BP layer, and (iii) resource sharing/time multiplexing of computing units in BP layers where parallel processing of the entire batch is neither necessary nor feasible in terms of resources. 
The compound effect of these optimizations in our BP HW implementation leads to a reduction of multiplication complexity by around $85$\%.  Although operator complexity is a blunt metric for predicting actual HW circuits, a significant complexity reduction is critical to achieving an ML equalizer implementation which can fit on an FPGA.

\section{Results}
\label{sec:results}

\begin{figure}[b]
    \vspace{-0.3cm}
    \centering
    \includegraphics[width=0.99\columnwidth]{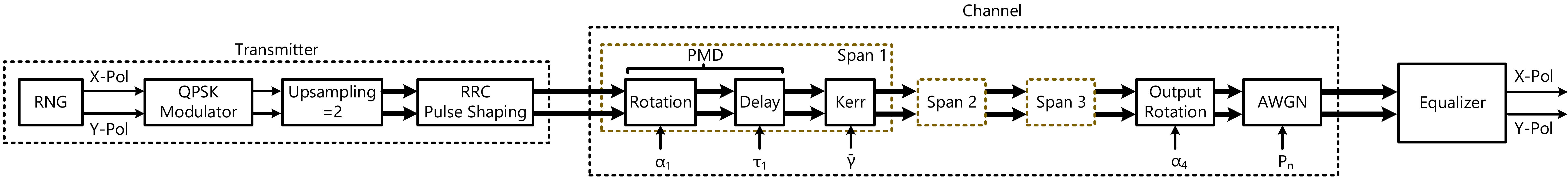}
    \vspace{-0.3cm}
    \caption{Block diagram of the end-to-end system model.}
    \label{fig:system_model}
    \vspace{-0.3cm}
\end{figure}

The system model (Fig.~\ref{fig:system_model}) is adapted from \cite{ModelBased} and intends to strike a balance between realism and the limited HW resources available for our equalizer, while still providing meaningful training data. 
The model emulates single-channel transmission of a $32$-Gbaud signal (PM-QPSK, root-raised cosine, roll-off 0.1) over $3$ fiber spans of length $L = 100$ km. 
Each span $k$ applies (i) polarization rotations $\left(\begin{smallmatrix} \cos \alpha_k & \sin \alpha_k \\ -\sin \alpha_k  
	& \cos \alpha_k
\end{smallmatrix}\right)$, where $\alpha_k \in [-\pi, \pi]$ is a random rotation angle, (ii) differential group delays according to $\text{diag}(e^{-\imag \omega \frac{\tau_k}{2}}, e^{\imag \omega \frac{\tau_k}{2}} )$, where $\tau_k = \tau \sqrt{3 \pi L/8}$ and $\tau = 0.2\,$ ps/$\sqrt{\text{km}}$, and (iii) Kerr nonlinearities 
$\boldsymbol{u} \exp(-\imag \frac{8}{9} \gamma L \|\boldsymbol{u}\|^2  )$, where $\gamma = 1.2$ rad/W/km. 
After applying a final output rotation, Gaussian noise with average noise power $P_n = -14$ dBm is added before low-pass filtering the received signal. 
The equalizer itself uses $5$-tap MIMO-FIR filters, leading to $20$ real-valued coefficients per step, i.e., $60$ real-valued trainable parameters in total. 
The learning rate $\xi$ is separately adjusted for each input power. 

\begin{figure}[t]
    \vspace{-0.3cm}
    \centering
    \hspace{-0.5cm}
    \includegraphics[width=0.36\columnwidth]{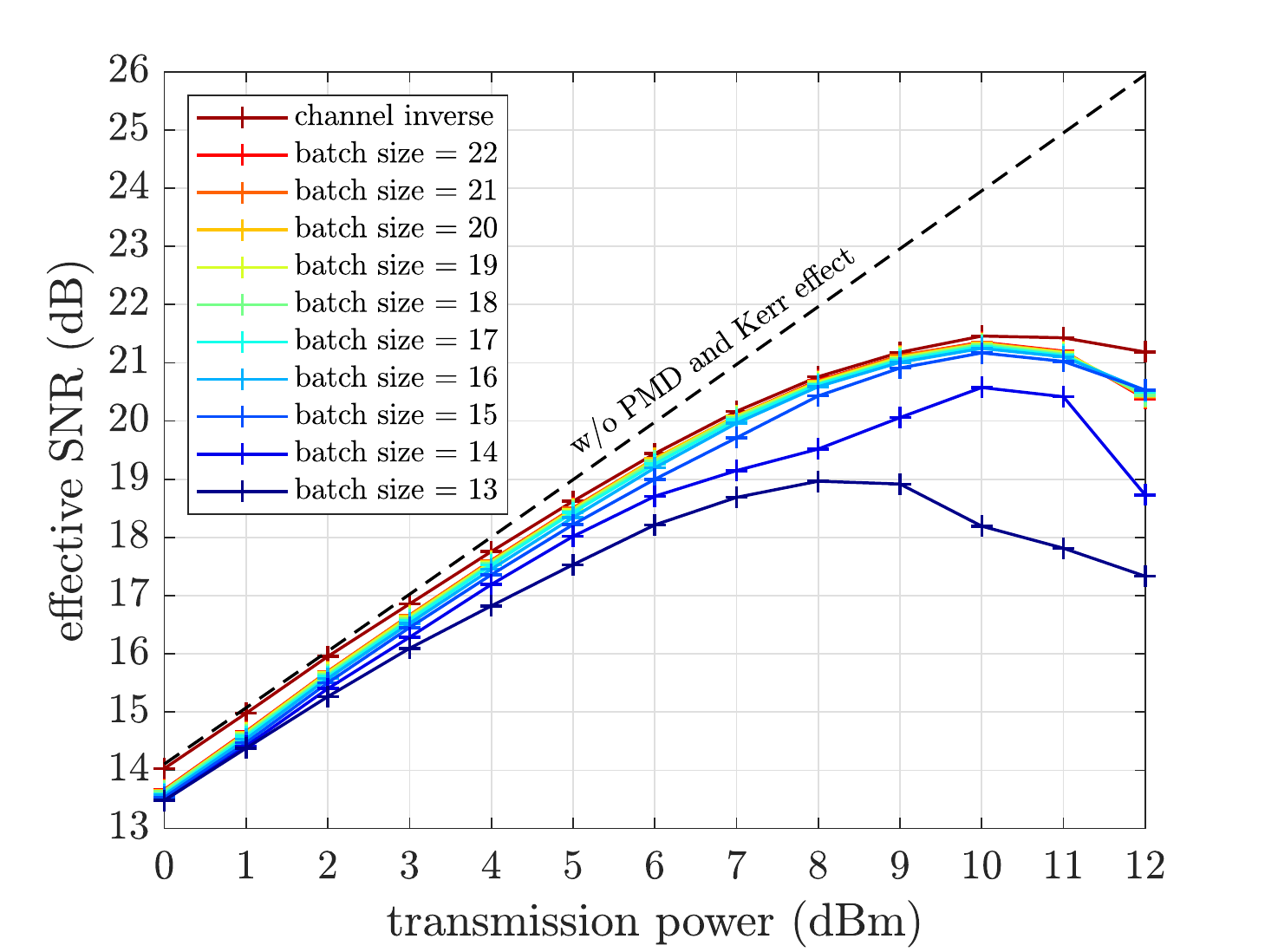}
    \hspace{-0.5cm}
    \includegraphics[width=0.36\columnwidth]{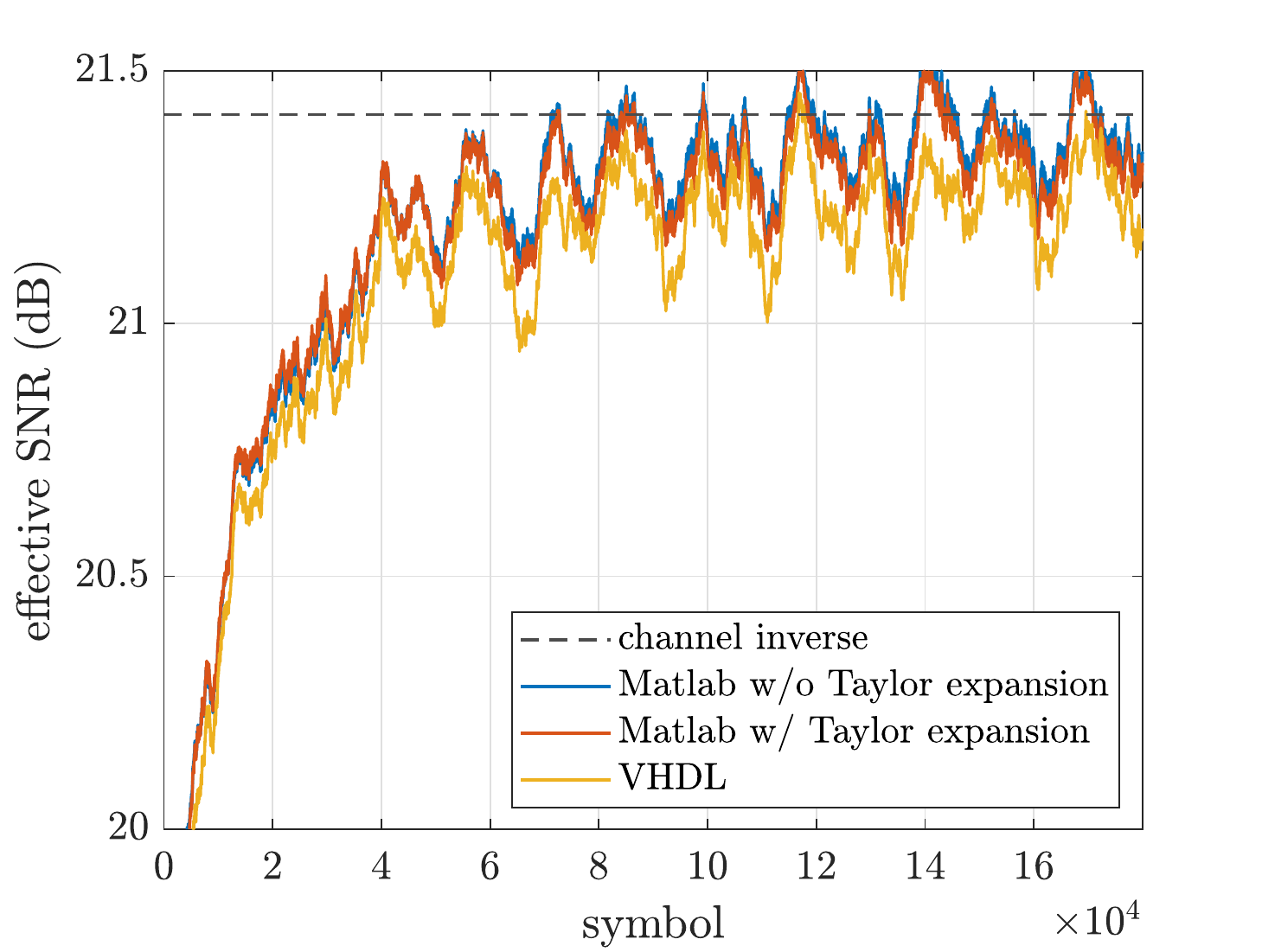}
    \hspace{-0.7cm}
    \includegraphics[width=0.36\columnwidth]{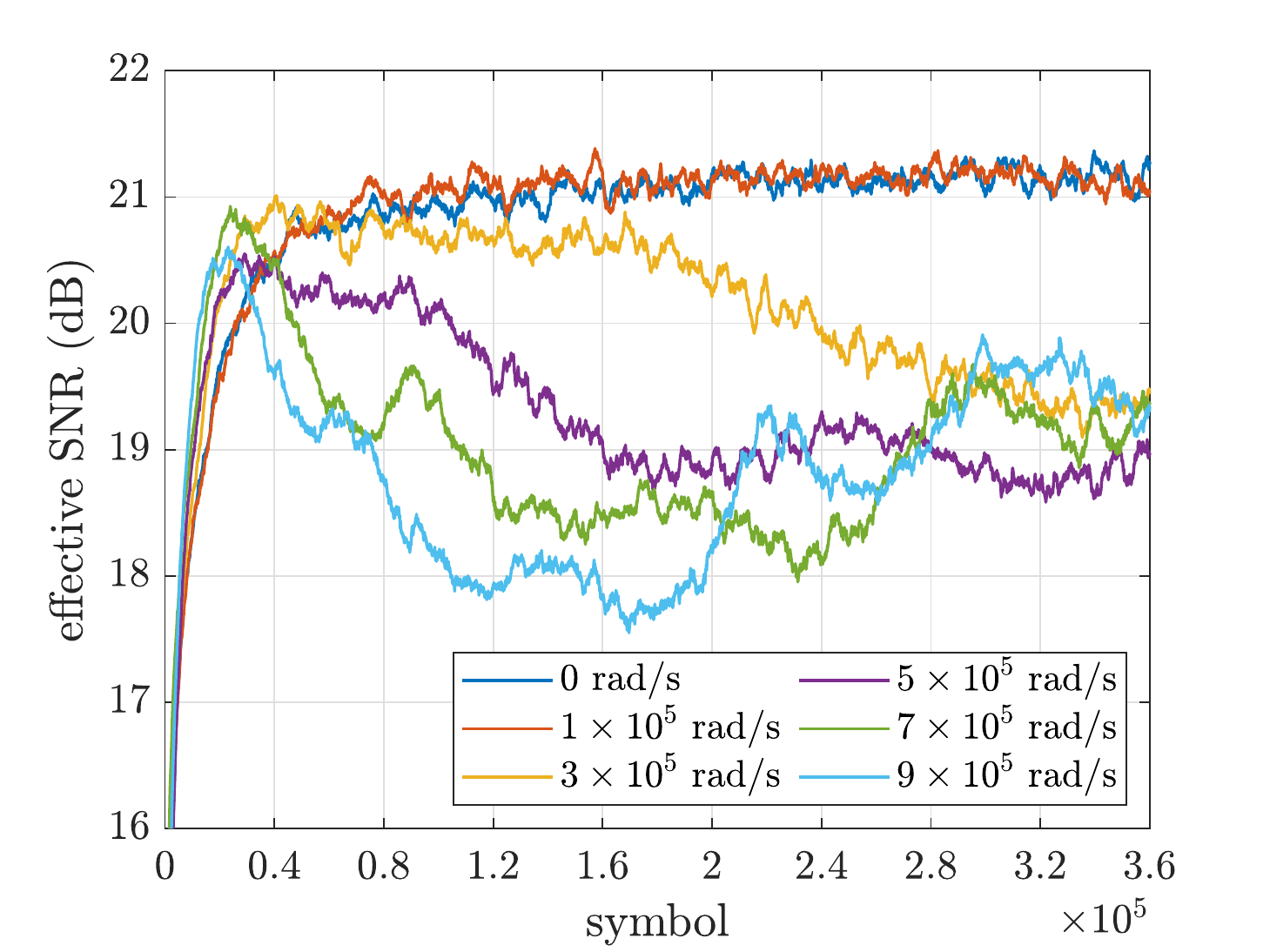}
    \hspace{-0.8cm}
    \vspace{-0.3cm}
    \caption{Results illustrating the overall system performance (left), convergence behavior (middle), and adaptivity (right). A sliding window averaging over $4096$ symbols is used in the middle and right figures. }
    \label{fig:results}
    \vspace{-0.5cm}
\end{figure}

\mysubsection{Batch size considerations}
Fig.~\ref{fig:results} (left) shows the system-level performance (in Matlab) in terms of the effective signal-to-noise ratio (SNR), i.e., the inverse of the MSE, as a function of the input power for various batch sizes. 
Larger batch sizes give better performance but are more difficult to implement in HW. 
For our implementation, we selected $B = 21$, which is close to the case where we initialize the equalizer to the (approximate) nonlinear channel inverse and comparable to, e.g., \cite{Crivelli2014} which uses $B=16$ for their adaptive linear equalizer. 

\mysubsection{Wordlength considerations}
Ideally, we want to reduce the signal wordlengths (WLs) so that we save HW resources but still have performance close to the system simulations. 
In the VHDL equalizer, we use five different WLs: (i) WL of symbol/sample
\textit{and} BP layers, except gradient layer output, (ii) WL of the parameter $\bar{\gamma} = \frac{8}{9} \gamma L$, (iii) WL of the Kerr angle $\phi = \bar{\gamma} \|\boldsymbol{u}\|^2$ in the Kerr FP layers, (iv) WL of each tap in the linear FP layers \textit{and} gradients in the gradient layers, and (v) WL of the taps in the MF layers. 
Using logic simulations, we assess the equalizer performance in terms of effective SNR and convergence speed. 
Our analysis shows that a good choice for the WLs are \{14,16,12,14,12\} which give a moderate penalty and a performance of around $21.2$ dB. 
By accepting some more performance degradations, WLs can be reduced further. 

\mysubsection{Convergence} Fig.~\ref{fig:results} (middle) compares the convergence behavior of the final VHDL equalizer at the optimal launch power of $10$ dBm to the reference Matlab implementation (using identical training data and noise realizations). 
While the Taylor expansion does not appreciably affect the effective SNR, the VHDL implementation is affected by fixed-point quantization and rounding and scaling
issues, leading to slightly worse performance.

\mysubsection{Time-varying channels and adaptivity}
Fig.~\ref{fig:results} (right) shows how the equalizer handles different 
polarization rotation speeds. 
Here, we assume that all four rotation angles $\alpha_k$ (3 spans + final rotation) are time-dependent, where the rotation speed is varied from $10^5$ to $10^6$ rad/s. 
It can be seen that the equalizer shows essentially no performance penalty for $10^5$ rad/s, whereas higher speeds lead to a performance degradation. 

\mysubsection{Resource utilization}
Using Vivado Design Suite 2020.2 with a clock rate of 50 MHz,
the equalizer is implemented on a Xilinx VC709
FPGA development board with a Virtex-7 XC7VX690T.  The resource
utilization is shown in Table~\ref{tab:utlizationSummary}.
DSP slices are the main resource bottleneck. 
However, while the equalizer uses 79.72\% of 
\begin{wraptable}[12]{r}{6.0cm}
\vspace*{-0.6cm}
\centering
\caption{Resource usage.}
\label{tab:utlizationSummary}
\vspace*{-0.2cm}
\centering
\footnotesize
\begin{tabular}{@{}c|c|c@{}}
\hline

Module     & LUTs        & DSP slices \\ \hline  \hline
Kerr FPs   & 11,685 (2.70\%)   & 78 (2.17\%) \\
Linear FPs & 1,476 (0.34\%)     & 288 (8.00\%) \\
MF FP      & 7,028 (1.62\%)    & 0 (0.00\%)    \\
Loss BP    & 2,920 (0.67\%)    & 0 (0.00\%)\\
MF BP      & 5,724 (1.32\%)    & 132 (3.67\%)\\
Linear BPs & 7,030 (1.62\%)    & 480 (13.33\%)\\
Kerr BPs   & 37,131 (8.57\%)   & 992 (27.56\%)\\
Gradients  & 44,917 (10.37\%)  & 900 (25.00\%)\\ \hline
FP         & 20,189 (4.66\%)   & 366 (10.17\%)\\
BP         & 97,722 (22.56\%)  & 2,504 (69.56\%)\\
Total      & 120,423 (27.80\%) & 2,870 (79.72\%)\\
\end{tabular}
\end{wraptable}
DSP slices, there is still enough room for further adding an on-board PMD-Kerr emulator~\cite{Liu2022sppcom}, enabling a complete real-time test platform.
Although the time-multiplexed computing units save considerable HW resources in, e.g., the linear BP layers, it is clear that the BP HW consumes more resources than the FP HW. 
This is not too surprising since even for adaptive linear equalizers, the HW overhead related to training is considerable \cite{Crivelli2014}.  
For future work, it might be interesting to reduce the number of trainable parameters by decomposing the full MIMO filters \cite{ModelBased, PMD} or to reduce the training overhead by applying ideas from meta learning \cite{Simeone2020}. 
Moreover, 
one could investigate the possibility of only updating a \emph{subset} of parameters in real-time, potentially also using a less frequent updating scheme, to reduce power dissipation~\cite{Fougstedt2016ofc}. 

\section{Conclusion}

Our work demonstrates feasibility of on-chip training of a multi-layer ML equalizer on an FPGA, which represents an important step towards efficient implementation on ASICs, where further wordlength optimizations can be carried out. 
By using comprehensive cycle-based logic simulations, we verified that our VHDL implementation realizes an ML-based adaptive equalizer which can operate in real-time. 
This new equalizer represents an important DSP enabler which can be integrated in our CHOICE environment~\cite{larssonedefors+:OFC2022, CHOICEcode, CHOICE}. 

\vspace{0.1cm}

\noindent\footnotesize{\textbf{Acknowledgements: }The work of C.~H\"ager was supported by the Swedish Research Council under grant no.~2020-04718.}

\vspace{-0.3cm}
\label{references}

\end{document}